\pdfoutput=1
\documentclass[twocolumn,aps,prd,showpacs,amsmath,amssymb,10pt,superscriptaddress]{revtex4-1}
\usepackage{amsmath}
\usepackage{cases}
\usepackage{amssymb}
\usepackage[dvips]{graphicx}
\usepackage{color}
\usepackage{multirow}
\usepackage{url}
\usepackage{hyperref}

\begin{document}

\title{Bimodal grain-size scaling of thermal transport in polycrystalline graphene from large-scale molecular dynamics simulations}

\author{Zheyong Fan}
\email[Corresponding author: ]{brucenju@gmail.com}
\affiliation
{COMP Centre of Excellence, Department of Applied Physics, Aalto University School of Science, P.O. Box 11000, FI-00076 Aalto, Espoo, Finland}
\author{Petri Hirvonen}
\affiliation
{COMP Centre of Excellence, Department of Applied Physics, Aalto University School of Science, P.O. Box 11000, FI-00076 Aalto, Espoo, Finland}
\author{Luiz Felipe C. Pereira}
\affiliation
{Departamento de F\'{\i}sica, Universidade Federal do Rio Grande do Norte, Natal, RN, 59078-900, Brazil}
\author{Mikko M. Ervasti}
\affiliation
{COMP Centre of Excellence, Department of Applied Physics, Aalto University School of Science, P.O. Box 11000, FI-00076 Aalto, Espoo, Finland}
\author{Ken R. Elder}
\affiliation
{Department of Physics, Oakland University, Rochester, Michigan 48309, USA}
\author{Davide Donadio}
\affiliation
{Department of Chemistry, University of California at Davis, One Shields Avenue, Davis, California 95616, USA}
\author{Ari Harju}
\affiliation
{COMP Centre of Excellence, Department of Applied Physics, Aalto University School of Science, P.O. Box 11000, FI-00076 Aalto, Espoo, Finland}
\author{Tapio Ala-Nissila}
\affiliation
{COMP Centre of Excellence, Department of Applied Physics, Aalto University School of Science, P.O. Box 11000, FI-00076 Aalto, Espoo, Finland}
\affiliation
{Department of Mathematical Sciences and Department of Physics, Loughborough University, Loughborough,
Leicestershire LE11 3TU, United Kingdom}

\date{\today}

\begin{abstract}

Grain boundaries in graphene are inherent in wafer-scale samples prepared by chemical vapor deposition. They can strongly influence the mechanical properties and electronic and heat transport in graphene. In this work, we employ extensive molecular dynamics simulations to study thermal transport in large suspended polycrystalline graphene samples. Samples of different controlled grain sizes are prepared by a recently developed efficient multiscale approach based on the phase field crystal model. In contrast to previous works, our results show that the scaling of the thermal conductivity with the grain size implies bimodal behaviour with two effective Kapitza lengths. The scaling is dominated by the out-of-plane (flexural) phonons with a Kapitza length that is an order of magnitude larger than that of the in-plane phonons. We also show that in order to get quantitative agreement with the most recent experiments, quantum corrections need to be applied  to both the Kapitza conductance of grain boundaries and the thermal conductivity of pristine graphene and the corresponding Kapitza lengths must be renormalized accordingly.
\end{abstract}

\maketitle

Chemical vapor deposition, currently the only practical approach to grow wafer-scale graphene necessary for industrial applications, produces polycrystalline graphene containing grain boundaries \cite{yazyev2014} acting as extended defects that may influence electrical and thermal transport \cite{cummings2014,isacsson2017}. The influence of grain boundaries on heat conduction in graphene has been theoretically studied using various  methods, including molecular dynamics (MD) simulations \cite{bagri2011,cao2012,helgee2014,mortazavi2014,liu2014,wang2014,hahn2016}, Landauer-B\"{u}ttiker formalism \cite{lu2012,serov2013}, and Boltzmann transport formalism \cite{aksamija2014}. Although it is well known that graphene samples prepared by chemical vapor deposition \cite{cai2010} have smaller thermal conductivity than those prepared by micromechanical exfoliation \cite{balandin2008}, experimental measurements of the Kapitza conductances of individual grain boundaries in bicrystalline graphene samples \cite{yasaei2015} and thermal conductivities of polycrystalline graphene samples with controlled grain sizes \cite{ma2017,lee2017} have only been attempted recently.

A central issue for polycrystalline samples is how the thermal conductivity scales with the grain size $d$. Previous MD studies have focused on individual grain boundaries \cite{bagri2011,cao2012,helgee2014} or have only considered relatively small grain sizes, typically of a few nanometers \cite{wang2014,mortazavi2014,liu2014,hahn2016}. However, due to the broad distribution of phonon mean free paths that extend well beyond one micron in pristine graphene, accurate determination of the scaling properties requires considering relatively large grains and sample sizes.

\begin{figure}[htb]
\begin{center}
\includegraphics[width=\columnwidth]{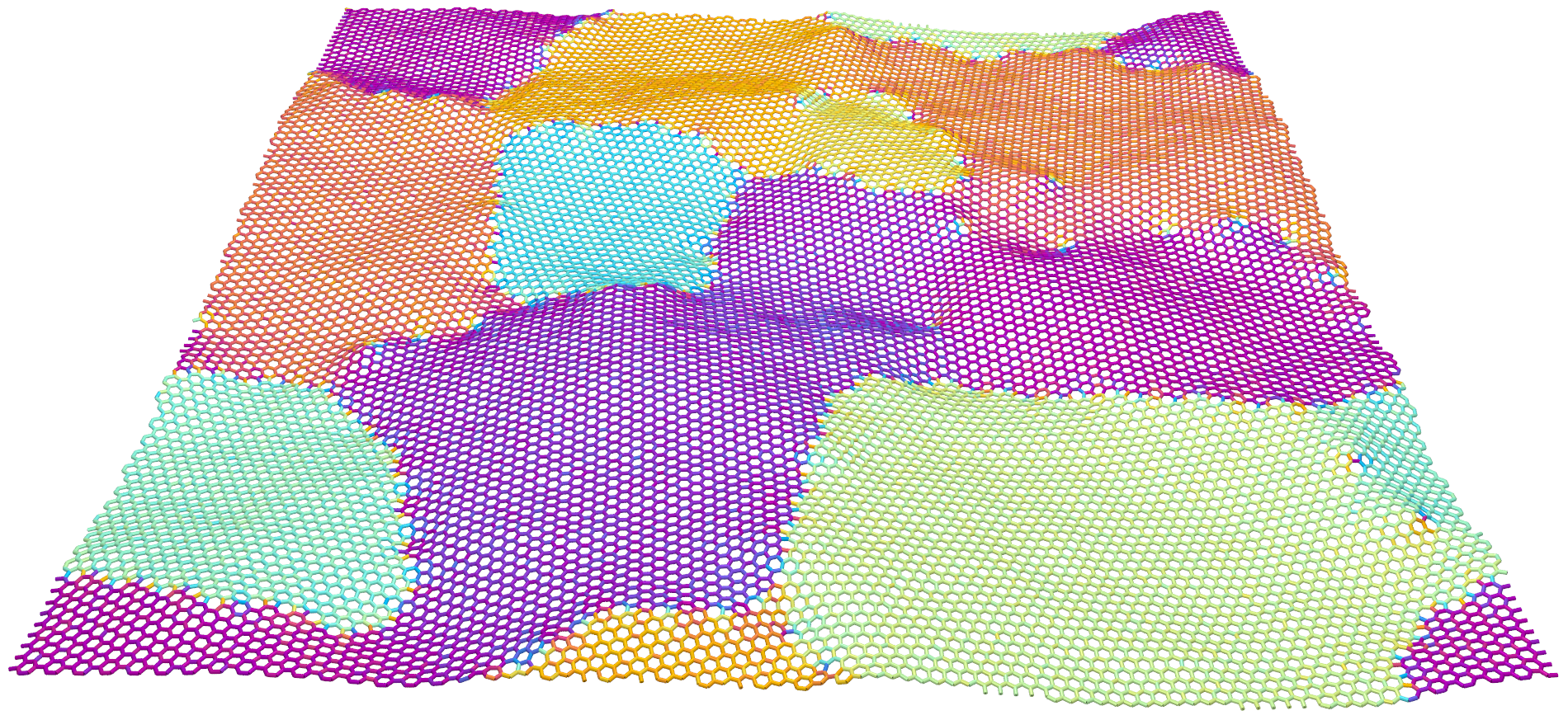}
\caption{A typical polycrystalline graphene sample after MD relaxation at $300$ K and zero in-plane pressure. This is the smallest sample size ($24 \times 24$ nm$^2$) considered here (see Methods for details). Different colors based on bond orientations are used to indicate different grains.}
\label{figure:sample}
\end{center}
\end{figure}

Recently, an efficient multiscale approach \cite{hirvonen2016,hirvonen2017} for modelling large polycrystalline graphene samples has been developed within the phase field crystal framework \cite{elder2002,elder2004,goldenfeld2005,mkhonta2013,zhang2014,seymour2016}. This method, combined with a newly developed highly efficient MD code for thermal conductivity calculations \cite{fan2013,fan2015,fan2017cpc,gpumd}, allows for direct atomistic simulations of the heat transport properties of large-scale realistic polycrystalline graphene samples (cf. Fig.  \ref{figure:sample}). In this Letter, we compare thermal conductivity values obtained by MD simulations from samples up to $192$ nm in linear size to recent experimental data \cite{ma2017}, resolving the discrepancy between the small Kapitza conductance ($3.8$ GW m$^{-2}$ K$^{-1}$) as extracted from the experimental data \cite{ma2017} and the larger values predicted from nonequilibrium MD \cite{bagri2011,cao2012} ($15 - 47$ GWm$^{-2}$K$^{-1}$) and Landauer-B\"{u}ttiker \cite{serov2013} ($\approx 8$ GWm$^{-2}$K$^{-1}$)  simulations. Further, we show that to obtain quantitative agreement with experiments quantum corrections need to be applied to classical simulation data.

\begin{figure}[htb]
\begin{center}
\includegraphics[width=\columnwidth]{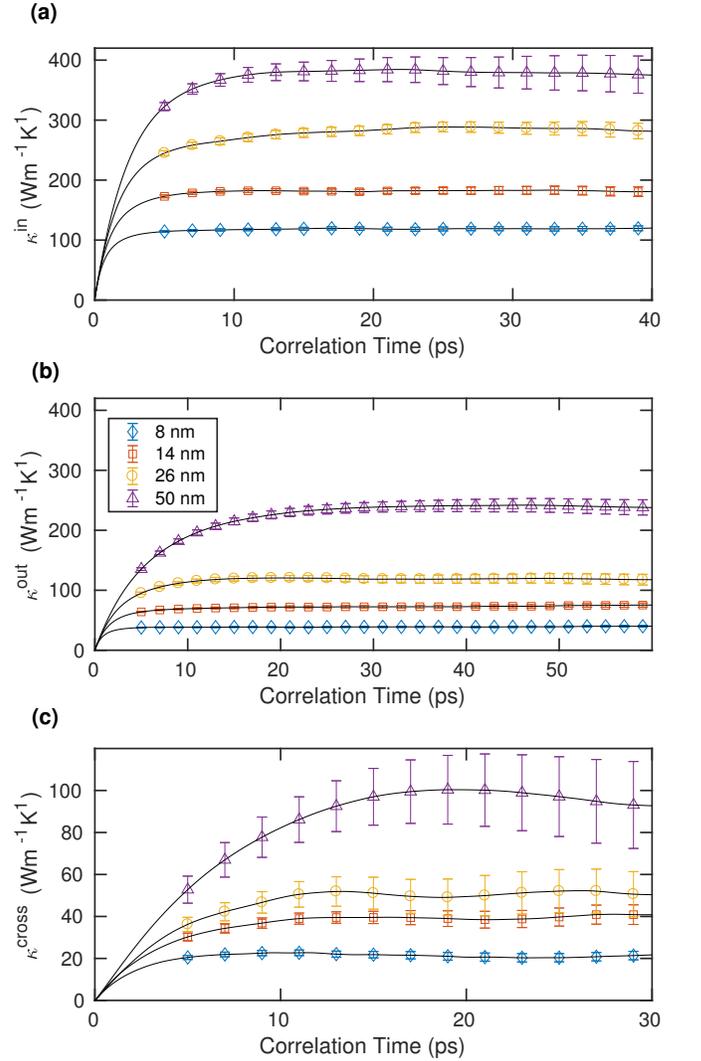}
\caption{Running thermal conductivity at $300$ K as a function of the correlation time for (a) the in-plane component, (b) the out-of-plane component, and (c) the cross term in polycrystalline graphene samples with different grain sizes $d = 8-50$ nm.}
\label{figure:size_rtc}
\end{center}
\end{figure}

We use the equilibrium Green-Kubo method \cite{green1954,kubo1957} within classical MD simulations to calculate the spatially resolved components \cite{fan2017prb}, $\kappa^{\text{in}}$ (for in-plane phonons), $\kappa^{\text{out}}$ (for out-of-plane, or flexural phonons), and $\kappa^{\text{cross}}$ (a cross term), of the thermal conductivity in polycrystalline graphene, as explained in Methods. Its running components with different grain sizes $d$ are shown in Fig. \ref{figure:size_rtc}. The first main result here is that time scale (which roughly corresponds to an average phonon relaxation time) for $\kappa^{\text{out}}$ is reduced from $\approx 1$ ns in pristine graphene (see Fig. 2 of Ref.~ \citenum{fan2017prb}) to $\approx 10$ ps in polycrystalline graphene. Thus in polycrystalline graphene $\kappa^{\text{out}} < \kappa^{\text{in}}$ in contrast to pristine graphene. This shows that the influence of grain boundaries is much stronger on the out-of-plane than the in-plane component. Another remarkable difference is that the cross term $\kappa^{\text{cross}}$ does not converge to zero as in pristine graphene, although its contribution is still relatively small. This is due to enhanced coupling between the out-of-plane and in-plane phonon modes in the presence of larger surface corrugation.

\begin{figure}[htb]
\begin{center}
\includegraphics[width=\columnwidth]{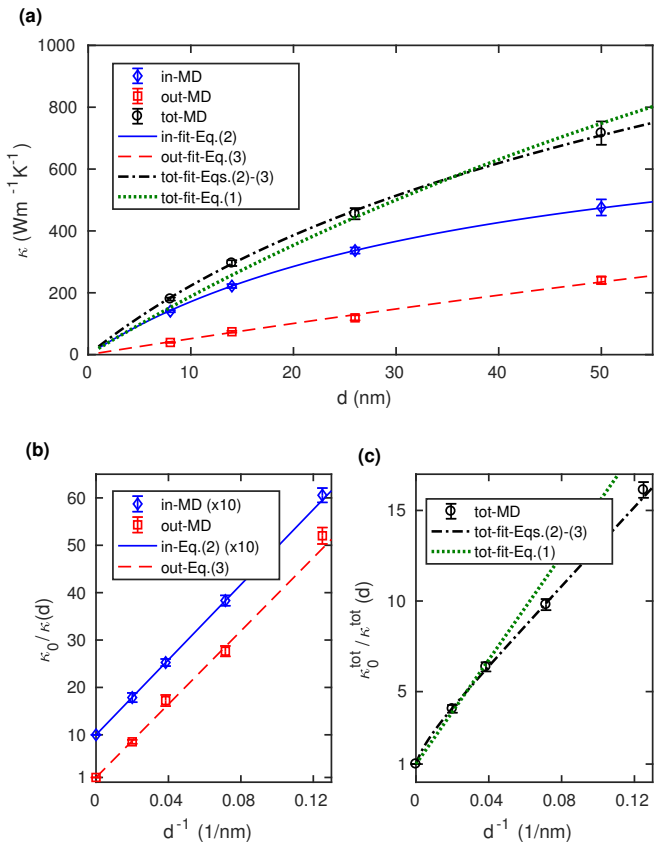}
\caption{(a) Thermal conductivity $\kappa(d)$ at $300$ K as a function of the effective grain size $d$ for the various components (in-plane, out-of-plane, and total). Markers are data from MD simulations and lines are fits. The green dotted line corresponds to Eq. (\ref{equation:scaling-1}) and the others to Eqs. (\ref{equation:scaling-2a}) - (\ref{equation:scaling-2b}). (b)-(c) Normalized inverse conductivity $\kappa_0/\kappa(d)$ as a function of inverse grain size $1/d$ for the various components. For clarity, the in-plane component $\kappa_0^{\text{in}}/\kappa^{\text{in}}(d)$ has been multiplied by a factor of ten. See text for details.}
\label{figure:size_emd_1}
\end{center}
\end{figure}

To study the scaling of $\kappa$ with $d$, we first plot $\kappa^{\text{in}}$, $\kappa^{\text{out}}$, and the total $\kappa^{\text{tot}} = \kappa^{\text{in}} + \kappa^{\text{out}}$ against $d$ in Fig. \ref{figure:size_emd_1} (we have included the small cross term $\kappa^{\text{cross}}$ into $\kappa^{\text{in}}$ as they have similar time scales). More details on the thermal conductivity data are presented in Supporting Information Figure S1 and Table S1. Previously, the scaling of $\kappa^{\text{tot}}$ with $d$ has been modelled by the following simple formula both in theoretical \cite{bagri2011,wang2014,hahn2016} and experimental works \cite{ma2017}:
\begin{equation}
1/\kappa^{\text{tot}}(d) = 1/\kappa^{\text{tot}}_0 + 1/(G^{\text{tot}}d),
\label{equation:scaling-1}
\end{equation}
where $\kappa^{\text{tot}}_0$ is the total thermal conductivity of pristine graphene (i.e., polycrystalline graphene in limit of infinite $d$) and $G^{\text{tot}}$ is the Kapitza conductance (or grain boundary conductance), which here characterises the average influence of the different grain boundaries on the heat flux across them.

In our previous MD simulations for pristine graphene \cite{fan2017prb} we have obtained $\kappa^{\text{tot}}_0 = 2~900 \pm 100$ Wm$^{-1}$K$^{-1}$. Thus the only unknown quantity in Eq. (\ref{equation:scaling-1}) is $G^{\text{tot}}$, which can be treated as a fitting parameter. In Fig. \ref{figure:size_emd_1}(a) it can been seen that the fit is not adequate. The reason is that the in-plane and out-of-plane components have very different properties, resulting in a nonlinear behavior of $1/\kappa^{\text{tot}}(d)$ with respect to $1/d$. Following Ref. \citenum{fan2017prb} we conclude that this bimodal behavior must be taken into account by separating the two components as
\begin{equation}
1/\kappa^{\text{in}}(d) = 1/\kappa^{\text{in}}_0 + 1/(G^{\text{in}}d);
\label{equation:scaling-2a}
\end{equation}
\begin{equation}
1/\kappa^{\text{out}}(d) = 1/\kappa^{\text{out}}_0 + 1/(G^{\text{out}}d),
\label{equation:scaling-2b}
\end{equation}
with $\kappa^{\text{tot}}(d) = \kappa^{\text{in}}(d) + \kappa^{\text{out}}(d)$, and $\kappa^{\text{in}}_0 \approx 850$ Wm$^{-1}$K$^{-1}$ and $\kappa^{\text{out}}_0 \approx 2~050$ Wm$^{-1}$K$^{-1}$ from Ref. \citenum{fan2017prb}. As it can be seen in Fig. \ref{figure:size_emd_1}, Eqs. (\ref{equation:scaling-2a}) - (\ref{equation:scaling-2b}) give accurate fits yielding $G^{\text{in}} \approx 21$ GWm$^{-2}$K$^{-1}$ and $G^{\text{out}}\approx 5$  GWm$^{-2}$K$^{-1}$. The total Kapitza conductance $G^{\text{tot}} = G^{\text{in}}+G^{\text{out}} \approx 26$ GWm$^{-2}$K$^{-1}$ lies well within the range predicted by nonequilibrium MD simulations \cite{bagri2011,cao2012}.

The scaling parameter $L^{i} = \kappa_0^{i}/G^{i} \quad (i=\text{in},~\text{out})$ in the equations above has the dimension of a length and it defines the Kapitza length \cite{nan1997}. In terms of the Kapitza length, the conductivity ratios $\kappa^{i}(d)/\kappa^{i}_0$  can be written as $\kappa^{i}(d)/\kappa^{i}_0 = 1/(1+L^{i}/d)$. This shows that when the grain size equals the Kapitza length, $\kappa^i(d)$ reaches half of $\kappa^i_0$. The Kapitza lengths for the in-plane and out-of-plane components from our MD data are $L^{\text{in}}\approx 40$ nm and $L^{\text{out}}\approx 400$ nm, differing by an order of magnitude which reflects the difference in the scaling of the corresponding conductivity components with $d$ (cf. Fig. \ref{figure:size_emd_1}).

\begin{figure}[htb]
\begin{center}
\includegraphics[width=\columnwidth]{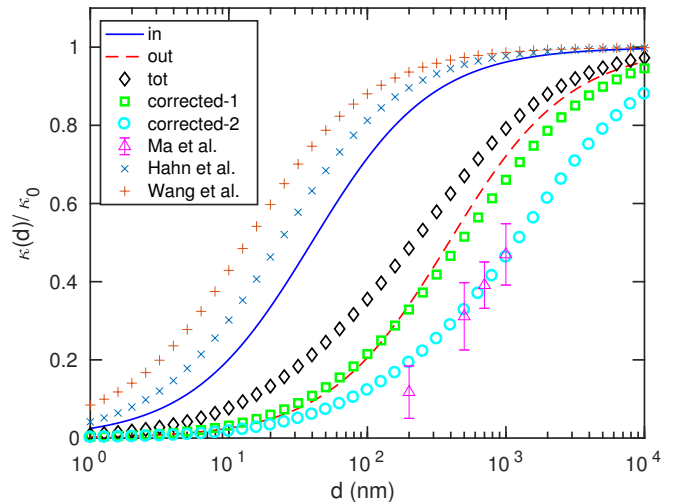}
\caption{Components of the normalized thermal conductivity $\kappa(d)/\kappa_{0}$ at $300$ K as a function of the effective grain size $d$. The solid and dashed lines represent the results for $\kappa^{\text{in}}(d)/\kappa^{\text{in}}_{0}$ and $\kappa^{\text{out}}(d)/\kappa^{\text{out}}_{0}$ as predicted by Eqs. (\ref{equation:scaling-2a}) and (\ref{equation:scaling-2b}), respectively. The diamonds, squares, and circles represent the results for $\kappa^{\text{tot}}(d)/\kappa^{\text{tot}}_{0}$ without quantum corrections, with the mode-to-mode quantum correction to $G$ (corrected-1), and with quantum corrections to both $G$ and $\kappa$ (corrected-2), respectively. The purple triangles are the experimental data from Ma \textit{et al.} \cite{ma2017}. The cross and plus symbols represent the theoretical predictions by Hahn \textit{et al.} \cite{hahn2016} and Wang \textit{et al.} \cite{wang2014}, respectively. See text for details.}
\label{figure:size_emd_2}
\end{center}
\end{figure}

In Fig. \ref{figure:size_emd_2} we compare our MD data for the scaling of the components of $\kappa(d)/\kappa_0$ vs. $d$ with previous theoretical predictions \cite{wang2014,hahn2016}. They are closer to our results for the in-plane component indicating that $\kappa^{\text{out}}(d)$ was not properly accounted for due to either an incorrect definition of the heat current or unconverged size scaling. In fact, the calculations in Ref.~ \citenum{wang2014} were based on the heat current formula in LAMMPS \cite{lammps} which is incorrect for many-body potentials \cite{fan2015}. Indeed, $\kappa^{\text{tot}}_0$ in Ref.~\citenum{wang2014} was estimated to be about 720 Wm$^{-1}$K$^{-1}$, which is even smaller than our $\kappa^{\text{in}}_0$. The calculations in Ref.~\citenum{hahn2016} were based on the approach-to-equilibrium MD method \cite{lampin2013,melis2014} with a fixed sample size of $200$ nm long and $20$ nm wide, which also significantly underestimates the contribution from the out-of-plane component. The Kapitza lengths for the data from Refs. \citenum{wang2014} and \citenum{hahn2016} can be estimated to be $\approx 15$ nm and $25$ nm, respectively, in stark contrast with our results.

\begin{figure}[htb]
\begin{center}
\includegraphics[width=\columnwidth]{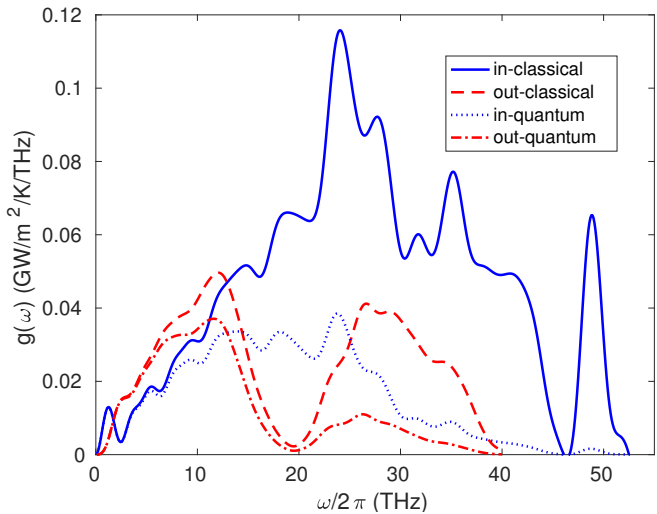}
\caption{Spectral conductance $g(\omega)$ in the $24\times 24$ nm$^2$ system at 300 K as a function of phonon frequency for the in-plane and out-of-plane components before (labeled as ``classical'') and after (labeled as ``quantum'') quantum correction.}
\label{figure:quantum}
\end{center}
\end{figure}

In Fig. \ref{figure:size_emd_2} we also plot the data from the most recent experimental measurements (triangles) \cite{ma2017}. Although our new data are much closer to the experiments than the previous theoretical results, there is still a quantitative difference. One reason for the discrepancy is the use of classical statistics in view of the high Debye temperature ($\approx 2000$ K) of graphene \cite{pop2012}. Using classical statistics can grossly overestimate the Kapitza conductance. Indeed, quantum mechanical calculations based on the Landauer-B\"{u}ttiker formalism \cite{lu2012,serov2013} predict graphene grain boundary conductance several times smaller than that from classical non-equilibrium MD simulations \cite{bagri2011,cao2012}. While there is no rigorous way to include all the quantum effects within the present calculations, we can gauge their importance by applying the mode-by-mode quantum correction in Ref.~\citenum{saaskilahti2016} to the spectral conductance. In Fig. \ref{figure:quantum} we show our data for the in-plane and out-of-plane components of the spectral conductance $g^{i}(\omega) ~ (i=\text{in},~\text{out})$ for the $24 \times 24$ nm$^2$ polycrystalline system, calculated using the spectral decomposition method in Ref.~\citenum{fan2017prb}.  The mode-to-mode quantum corrections can be incorporated by multiplying $g^{i}(\omega) ~ (i=\text{in},~\text{out})$ by the factor $x^2e^x/(e^x-1)^2$ ($x = \hbar \omega / k_B T$), which yields the dotted and dot-dashed lines in Fig. \ref{figure:quantum}. The influence of these corrections is significant; the integrated conductance is reduced by a factor of three for the in-plane component and by a factor of two for the out-of-plane component. Therefore, our estimates for $G^{\text{in}}$ and $G^{\text{out}}$ are modified to $7$ GWm$^{-2}$K$^{-1}$ and $2.5$ GWm$^{-2}$K$^{-1}$, respectively. The corresponding Kapitza lengths are changed to $L^{\text{in}}\approx 120$ nm and $L^{\text{out}}\approx 800$ nm, respectively. The results for the modified conductivity ratio $\kappa^{\text{tot}}(d)/\kappa^{\text{tot}}_0$ are plotted in Fig. \ref{figure:size_emd_2} with squares. The agreement with the experiments is better, but still not at a quantitative level.

To fully resolve the discrepancy between our data and the experiments we need to revisit the case of pristine graphene. Classical statistics can underestimate the thermal conductivity in the pristine case, too, by overestimating the phonon-phonon scattering rates of the low-frequency phonon modes \cite{turney2009,singh2011}, the major heat carriers in pristine graphene. This explains why the thermal conductivity of pristine graphene calculated by our previous MD simulations ($2~900$ Wm$^{-1}$K$^{-1}$) is significantly smaller than the prediction from lattice dynamics calculations \cite{lindsay2010prb,aksamija2011apl,singh2011,majee2016prb,kuang2016} (about $5~000$ Wm$^{-1}$K$^{-1}$)  and the most recent experiments on high-quality monocrystalline graphene \cite{ma2017} ($5~200$ Wm$^{-1}$K$^{-1}$) . Unfortunately, unlike in the case of grain boundary conductance, there is so far no feasible quantum correction method for classical MD thermal conductivity calculations in the diffusive regime where phonon-phonon scattering dominates. In view of the fact that the differences between the results from classical MD and quantum mechanical lattice dynamics methods mainly concern the out-of-plane component \cite{fan2017prb,kuang2016}, we can resolve this issue here by scaling $\kappa_0^{\text{out}}$ such that $\kappa_0^{\text{tot}}$ equals the experimental reference value \cite{ma2017} of $5~200$ Wm$^{-1}$K$^{-1}$. Combining this with the mode-to-mode quantum corrections to the Kapitza conductance above, the final revised Kapitza lengths for the two components now become $L^{\text{in}}\approx 0.12$ $\mu$m and $L^{\text{out}}\approx 2$ $\mu$m, respectively, differing by more than an order of magnitude. With quantum corrections to both $G$ and $\kappa$, the scaling of $\kappa(d)/\kappa_0$ (circles in Fig. \ref{figure:size_emd_2}) finally agrees with the experiments at a fully quantitative level.

Finally, we note that in the experimental work \cite{ma2017}, the grain-size scaling was interpreted in terms of a single Kapitza conductance of $3.8$ GWm$^{-2}$K$^{-1}$, which is much smaller than that from Landauer-B\"{u}ttiker calculations \cite{serov2013} ($8$ GWm$^{-2}$K$^{-1}$). In contrast, our bimodal grain size scaling with the two quantum corrections gives $G^{\text{tot}} = G^{\text{in}} + G^{\text{out}}=9.5$ GWm$^{-2}$K$^{-1}$, which is a reasonable value considering that the harmonic approximation used in Ref.~\citenum{serov2013} can somewhat underestimate the Kapitza conductance at room temperature. 

In summary, by using high-accuracy MD simulations of large polycrystalline graphene samples generated by a multiscale modeling approach, we have demonstrated that the inverse thermal conductivity does not scale linearly with respect to the inverse grain size but shows a bimodal behavior. The Kapitza lengths for the grain boundaries associated with the in-plane and out-of-plane phonon branches differ by more than one order of magnitude. While the grain-size scaling is dominated by the out-of-plane phonons with a much larger Kapitza length, the in-plane phonons contribute more to the Kapitza conductance. We have also demonstrated that in order to obtain quantitative agreement with the most recent experiments of heat conduction in polycrystalline and pristine graphene samples, quantum corrections to both the Kapitza conductance of grain boundaries and the thermal conductivity of pristine graphene must be included and the corresponding Kapitza lengths must be renormalized accordingly. 

We note that we have only considered suspended graphene samples in this work. For supported graphene, heat transport by the out-of-plane phonons will be significantly suppressed. Whether or not the bimodal scaling will survive in the presence of a substrate is an interesting question which requires further study. Finally, we point out that our samples were generated by the phase field crystal method, which has been shown to reproduce realistic grain size distributions in the asymptotic limit in two dimensions \cite{backofen2014}. Such samples may not correspond to those observed in some experiments \cite{lee2017}, but additional non-uniformity and anisotropy should influence the results only quantitatively, not qualitatively. The concept of an effective grain size is analogous to that of an effective phonon mean free path, which, in spite of being a relatively crude estimate, captures the essential physics. Because the in-plane and out-of-plane phonons have drastically distinct transport properties, we expect that the bimodal scaling would survive even if the influence of additional non-uniformity and anisotropy were taken into account. This argument is further supported by the fact that while the Kapitza conductance of individual grain boundaries depends on the angle of misorientation, for angles between about 20 and 40 degrees, this dependence is rather weak. We have carried out a comprehensive study of the Kapitza conductance for grain boundaries of different orientations and the results will be published elsewhere.

{\bf Methods.} Realistic polycrystalline samples are constructed with the phase field crystal model by starting with small random crystallites that grow in a disordered density field. We grow the polycrystalline samples akin to chemical vapor deposition by assuming non-conserved dynamics. The relaxed density field is converted into a discrete set of atomic coordinates suited for the initialization of MD simulations \cite{hirvonen2016,hirvonen2017}. To investigate the scaling properties, we construct polycrystalline samples with various characteristic grain sizes, $d$, defined as $d=(A/n)^{1/2}$, where $A$ is the total planar area, and $n$ is the number of grains comprising it. We consider  systems of four sizes: $24 \times 24$ nm$^2$, $48 \times 48$ nm$^2$, $96 \times 96$ nm$^2$, and $192 \times 192$ nm$^2$. Each case was initialized  with 16 randomly placed and oriented crystallites. The final number of grains in a sample is typically smaller than 16 and the effective grain sizes, averaged over a few realizations for each sample size, are found to be 8 nm, 14 nm, 26 nm, and 50 nm, respectively. Figure \ref{figure:sample} shows a typical structure of our smallest $24 \times 24$ nm$^2$ polycrystalline sample after MD relaxation.

We use the Green-Kubo method \cite{green1954,kubo1957}, with equilibrium MD, to calculate the thermal conductivities. In this method, one can calculate the running thermal conductivity tensor  $\kappa_{\mu\nu}(t)$ ($\mu$ and  $\nu$ can be $x$ or $y$ for two-dimensional materials) as a function of the correlation time $t$ as $\kappa_{\mu\nu}(t) = \int_0^{t} \langle J_{\mu}(0)J_{\nu}(t') \rangle dt'$, where $\langle J_{\mu}(0)J_{\nu}(t') \rangle$ is the heat current autocorrelation function evaluated as the time average $\langle~ \rangle$ of the product of two heat currents separated by $t'$. A decomposition of the heat current, $J_{\mu}=J_{\mu}^{\text{in}}+J_{\mu}^{\text{out}}$, which is essential for two-dimensional materials, was introduced in Ref.~\citenum{fan2017prb}. We stress that the decomposition is in terms of the velocity, and the out-of-plane heat current $J_{\mu}^{\text{out}}$ is not a heat current perpendicular to the graphene basal plane (taken as the $xy$ plane), but a heat current in the basal plane ($\mu$ direction) contributed by the out-of-plane vibrational modes.  With this decomposition, the thermal conductivity decomposes into three terms, $\kappa^{\text{in}}_{\mu\nu}(t)$, $\kappa^{\text{out}}_{\mu\nu}(t)$, and $\kappa^{\text{cross}}_{\mu\nu}(t)$, associated with the autocorrelation functions $\langle J^{\text{in}}_{\mu}(0)J^{\text{in}}_{\nu}(t')\rangle$, $\langle J^{\text{out}}_{\mu}(0)J^{\text{out}}_{\nu}(t')\rangle$, and $2\langle J^{\text{in}}_{\mu}(0)J^{\text{out}}_{\nu}(t')\rangle$, respectively. As our systems can be assumed statistically isotropic, the conductivity in the basal plane can be treated as a scalar $\kappa = (\kappa_{xx}+\kappa_{yy})/2$.

We perform the equilibrium MD simulations using an efficient GPUMD code \cite{fan2013,fan2015,fan2017cpc}. The Tersoff potential \cite{tersoff1989} optimized for graphene \cite{lindsay2010} is used here. The velocity-Verlet method \cite{swope1982} is used for time integration, with a time step of $1$ fs for all the systems. All the simulations are performed at $300$ K and a weak coupling thermostat (Berendsen) \cite{berendsen1984} is used to control temperature and pressure during the equilibration stage (which lasts $2$ ns for all the systems). The production stage (in which heat current is recorded) lasts $5$ ns and $10$ independent simulations are performed for each sample. Periodic boundary conditions are applied on the $xy$ plane.  Although the magnitude of the out-of-plane deformation can exceed $1$ nm in some cases, we assume a uniform thickness of $0.335$ nm for the monolayer when reporting the effective three-dimensional thermal conductivity values.

The spectral conductance $g^{i}(\omega) ~ (i=\text{in},~\text{out})$ is calculated using a spectral decomposition method \cite{Saaskilahti2014,zhou2015,fan2017prb} in the framework of nonequilibrium MD simulations. The system is divided into a number of blocks along the transport direction, with the two outermost blocks being taken as heat source and sink, maintained at 320 K and 280 K, respectively, using the Nos\'{e}-Hoover chain thermostat \cite{nose1984,hoover1985,martyna1992}. The transverse direction is treated as periodic and the two edges in the transport direction are fixed. After achieving steady state, we calculate the correlation function $K^{i}_{A \rightarrow B}(t) ~ (i=\text{in},~\text{out})$ defined in Ref.~\citenum{fan2017prb}. Then the spectral conductance is calculated as \cite{fan2017prb} $g^{i}(\omega) =  \int_{-\infty}^{+\infty} dt e^{i\omega t} \left[ 2K^{i}_{A \rightarrow B}(t)/(S\Delta T) \right]$, where $S$ is the cross-sectional area and $\Delta T$ is the temperature difference between the source and the sink.

\begin{figure}[htb]
\begin{center}
\includegraphics[width=\columnwidth]{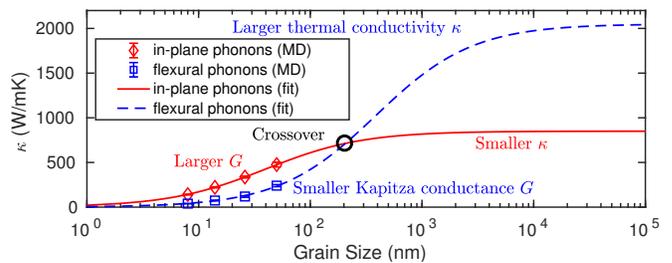}
\caption{The TOC figure.}
\end{center}
\end{figure}

\noindent 
\textbf{Supporting information}

The following files are available  free of charge.
\begin{itemize}
  \item samples.zip: All the polycrystalline graphene samples created by the phase field crystal method.
  \item supp.pdf: Detailed results on the thermal conductivity of the polycrystalline graphene samples.
\end{itemize}

\noindent
\textbf{Notes}

The authors declare no competing financial interests.

\begin{acknowledgments}

This research has been supported in small part by the Academy of Finland through its Centres of Excellence Program (Project No. 251748). We acknowledge the computational resources provided by Aalto Science-IT project and Finland's IT Center for Science (CSC). Z.F. acknowledges the support from the National Natural Science Foundation of China (Grant No. 11404033). P.H. acknowledges financial support from the Foundation for Aalto University Science and Technology, and from the Vilho, Yrj\"o and Kalle V\"ais\"al\"a Foundation of the Finnish Academy of Science and Letters. L.F.C.P. acknowledges financial support from the Brazilian government agency CAPES for project ``Physical properties of nanostructured materials'' (Grant No. 3195/2014) via its Science Without Borders program. K.R.E. acknowledges financial support from the US National Science Foundation under Grant No. DMR-1506634.

\end{acknowledgments}

\bibliography{poly}

\begin{thebibliography}{55}%
\makeatletter
\providecommand \@ifxundefined [1]{%
 \@ifx{#1\undefined}
}%
\providecommand \@ifnum [1]{%
 \ifnum #1\expandafter \@firstoftwo
 \else \expandafter \@secondoftwo
 \fi
}%
\providecommand \@ifx [1]{%
 \ifx #1\expandafter \@firstoftwo
 \else \expandafter \@secondoftwo
 \fi
}%
\providecommand \natexlab [1]{#1}%
\providecommand \enquote  [1]{``#1''}%
\providecommand \bibnamefont  [1]{#1}%
\providecommand \bibfnamefont [1]{#1}%
\providecommand \citenamefont [1]{#1}%
\providecommand \href@noop [0]{\@secondoftwo}%
\providecommand \href [0]{\begingroup \@sanitize@url \@href}%
\providecommand \@href[1]{\@@startlink{#1}\@@href}%
\providecommand \@@href[1]{\endgroup#1\@@endlink}%
\providecommand \@sanitize@url [0]{\catcode `\\12\catcode `\$12\catcode
  `\&12\catcode `\#12\catcode `\^12\catcode `\_12\catcode `\%12\relax}%
\providecommand \@@startlink[1]{}%
\providecommand \@@endlink[0]{}%
\providecommand \url  [0]{\begingroup\@sanitize@url \@url }%
\providecommand \@url [1]{\endgroup\@href {#1}{\urlprefix }}%
\providecommand \urlprefix  [0]{URL }%
\providecommand \Eprint [0]{\href }%
\providecommand \doibase [0]{http://dx.doi.org/}%
\providecommand \selectlanguage [0]{\@gobble}%
\providecommand \bibinfo  [0]{\@secondoftwo}%
\providecommand \bibfield  [0]{\@secondoftwo}%
\providecommand \translation [1]{[#1]}%
\providecommand \BibitemOpen [0]{}%
\providecommand \bibitemStop [0]{}%
\providecommand \bibitemNoStop [0]{.\EOS\space}%
\providecommand \EOS [0]{\spacefactor3000\relax}%
\providecommand \BibitemShut  [1]{\csname bibitem#1\endcsname}%
\let\auto@bib@innerbib\@empty
\bibitem [{\citenamefont {Yazyev}\ and\ \citenamefont
  {Chen}(2014)}]{yazyev2014}%
  \BibitemOpen
  \bibfield  {author} {\bibinfo {author} {\bibfnamefont {O.~V.}\ \bibnamefont
  {Yazyev}}\ and\ \bibinfo {author} {\bibfnamefont {Y.~P.}\ \bibnamefont
  {Chen}},\ }\href@noop {} {\bibfield  {journal} {\bibinfo  {journal} {Nat.
  Nanotech.}\ }\textbf {\bibinfo {volume} {9}},\ \bibinfo {pages} {755}
  (\bibinfo {year} {2014})}\BibitemShut {NoStop}%
\bibitem [{\citenamefont {Cummings}\ \emph {et~al.}(2014)\citenamefont
  {Cummings}, \citenamefont {Duong}, \citenamefont {Nguyen}, \citenamefont
  {Tuan}, \citenamefont {Kotakoski}, \citenamefont {Vargas}, \citenamefont
  {Lee},\ and\ \citenamefont {Roche}}]{cummings2014}%
  \BibitemOpen
  \bibfield  {author} {\bibinfo {author} {\bibfnamefont {A.~W.}\ \bibnamefont
  {Cummings}}, \bibinfo {author} {\bibfnamefont {D.~L.}\ \bibnamefont {Duong}},
  \bibinfo {author} {\bibfnamefont {V.~L.}\ \bibnamefont {Nguyen}}, \bibinfo
  {author} {\bibfnamefont {D.~V.}\ \bibnamefont {Tuan}}, \bibinfo {author}
  {\bibfnamefont {J.}~\bibnamefont {Kotakoski}}, \bibinfo {author}
  {\bibfnamefont {J.~E.~B.}\ \bibnamefont {Vargas}}, \bibinfo {author}
  {\bibfnamefont {Y.~H.}\ \bibnamefont {Lee}}, \ and\ \bibinfo {author}
  {\bibfnamefont {S.}~\bibnamefont {Roche}},\ }\href@noop {} {\bibfield
  {journal} {\bibinfo  {journal} {Adv. Mater.}\ }\textbf {\bibinfo {volume}
  {26}},\ \bibinfo {pages} {5079} (\bibinfo {year} {2014})}\BibitemShut
  {NoStop}%
\bibitem [{\citenamefont {Isacsson}\ \emph {et~al.}(2017)\citenamefont
  {Isacsson}, \citenamefont {Cummings}, \citenamefont {Colombo}, \citenamefont
  {Colombo}, \citenamefont {Kinaret},\ and\ \citenamefont
  {Roche}}]{isacsson2017}%
  \BibitemOpen
  \bibfield  {author} {\bibinfo {author} {\bibfnamefont {A.}~\bibnamefont
  {Isacsson}}, \bibinfo {author} {\bibfnamefont {A.~W.}\ \bibnamefont
  {Cummings}}, \bibinfo {author} {\bibfnamefont {L.}~\bibnamefont {Colombo}},
  \bibinfo {author} {\bibfnamefont {L.}~\bibnamefont {Colombo}}, \bibinfo
  {author} {\bibfnamefont {J.~M.}\ \bibnamefont {Kinaret}}, \ and\ \bibinfo
  {author} {\bibfnamefont {S.}~\bibnamefont {Roche}},\ }\href@noop {}
  {\bibfield  {journal} {\bibinfo  {journal} {2D Mater.}\ }\textbf {\bibinfo
  {volume} {4}},\ \bibinfo {pages} {012002} (\bibinfo {year}
  {2017})}\BibitemShut {NoStop}%
\bibitem [{\citenamefont {Bagri}\ \emph {et~al.}(2011)\citenamefont {Bagri},
  \citenamefont {Kim}, \citenamefont {Ruoff},\ and\ \citenamefont
  {Shenoy}}]{bagri2011}%
  \BibitemOpen
  \bibfield  {author} {\bibinfo {author} {\bibfnamefont {A.}~\bibnamefont
  {Bagri}}, \bibinfo {author} {\bibfnamefont {S.-P.}\ \bibnamefont {Kim}},
  \bibinfo {author} {\bibfnamefont {R.~S.}\ \bibnamefont {Ruoff}}, \ and\
  \bibinfo {author} {\bibfnamefont {V.~B.}\ \bibnamefont {Shenoy}},\
  }\href@noop {} {\bibfield  {journal} {\bibinfo  {journal} {Nano Lett.}\
  }\textbf {\bibinfo {volume} {11}},\ \bibinfo {pages} {3917} (\bibinfo {year}
  {2011})}\BibitemShut {NoStop}%
\bibitem [{\citenamefont {Cao}\ and\ \citenamefont {Qu}(2012)}]{cao2012}%
  \BibitemOpen
  \bibfield  {author} {\bibinfo {author} {\bibfnamefont {A.}~\bibnamefont
  {Cao}}\ and\ \bibinfo {author} {\bibfnamefont {J.}~\bibnamefont {Qu}},\
  }\href@noop {} {\bibfield  {journal} {\bibinfo  {journal} {J. Appl. Phys.}\
  }\textbf {\bibinfo {volume} {111}},\ \bibinfo {pages} {053529} (\bibinfo
  {year} {2012})}\BibitemShut {NoStop}%
\bibitem [{\citenamefont {Helgee}\ and\ \citenamefont
  {Isacsson}(2014)}]{helgee2014}%
  \BibitemOpen
  \bibfield  {author} {\bibinfo {author} {\bibfnamefont {E.~E.}\ \bibnamefont
  {Helgee}}\ and\ \bibinfo {author} {\bibfnamefont {A.}~\bibnamefont
  {Isacsson}},\ }\href@noop {} {\bibfield  {journal} {\bibinfo  {journal}
  {Phys. Rev. B}\ }\textbf {\bibinfo {volume} {90}},\ \bibinfo {pages} {045416}
  (\bibinfo {year} {2014})}\BibitemShut {NoStop}%
\bibitem [{\citenamefont {Mortazavi}\ \emph {et~al.}(2014)\citenamefont
  {Mortazavi}, \citenamefont {P\"{o}tschke},\ and\ \citenamefont
  {Cuniberti}}]{mortazavi2014}%
  \BibitemOpen
  \bibfield  {author} {\bibinfo {author} {\bibfnamefont {B.}~\bibnamefont
  {Mortazavi}}, \bibinfo {author} {\bibfnamefont {M.}~\bibnamefont
  {P\"{o}tschke}}, \ and\ \bibinfo {author} {\bibfnamefont {G.}~\bibnamefont
  {Cuniberti}},\ }\href@noop {} {\bibfield  {journal} {\bibinfo  {journal}
  {Nanoscale}\ }\textbf {\bibinfo {volume} {6}},\ \bibinfo {pages} {3344}
  (\bibinfo {year} {2014})}\BibitemShut {NoStop}%
\bibitem [{\citenamefont {Liu}\ \emph {et~al.}(2014)\citenamefont {Liu},
  \citenamefont {Lin},\ and\ \citenamefont {Luo}}]{liu2014}%
  \BibitemOpen
  \bibfield  {author} {\bibinfo {author} {\bibfnamefont {H.~K.}\ \bibnamefont
  {Liu}}, \bibinfo {author} {\bibfnamefont {Y.}~\bibnamefont {Lin}}, \ and\
  \bibinfo {author} {\bibfnamefont {S.~N.}\ \bibnamefont {Luo}},\ }\href@noop
  {} {\bibfield  {journal} {\bibinfo  {journal} {J. Phys. Chem. C}\ }\textbf
  {\bibinfo {volume} {118}},\ \bibinfo {pages} {24797} (\bibinfo {year}
  {2014})}\BibitemShut {NoStop}%
\bibitem [{\citenamefont {Wang}\ \emph {et~al.}(2014)\citenamefont {Wang},
  \citenamefont {Song},\ and\ \citenamefont {Xu}}]{wang2014}%
  \BibitemOpen
  \bibfield  {author} {\bibinfo {author} {\bibfnamefont {Y.}~\bibnamefont
  {Wang}}, \bibinfo {author} {\bibfnamefont {Z.}~\bibnamefont {Song}}, \ and\
  \bibinfo {author} {\bibfnamefont {Z.}~\bibnamefont {Xu}},\ }\href@noop {}
  {\bibfield  {journal} {\bibinfo  {journal} {J. Mater. Res.}\ }\textbf
  {\bibinfo {volume} {29}},\ \bibinfo {pages} {362} (\bibinfo {year}
  {2014})}\BibitemShut {NoStop}%
\bibitem [{\citenamefont {Hahn}\ \emph {et~al.}(2016)\citenamefont {Hahn},
  \citenamefont {Melis},\ and\ \citenamefont {Colombo}}]{hahn2016}%
  \BibitemOpen
  \bibfield  {author} {\bibinfo {author} {\bibfnamefont {K.~R.}\ \bibnamefont
  {Hahn}}, \bibinfo {author} {\bibfnamefont {C.}~\bibnamefont {Melis}}, \ and\
  \bibinfo {author} {\bibfnamefont {L.}~\bibnamefont {Colombo}},\ }\href@noop
  {} {\bibfield  {journal} {\bibinfo  {journal} {Carbon}\ }\textbf {\bibinfo
  {volume} {96}},\ \bibinfo {pages} {429} (\bibinfo {year} {2016})}\BibitemShut
  {NoStop}%
\bibitem [{\citenamefont {Lu}\ and\ \citenamefont {Gao}(2012)}]{lu2012}%
  \BibitemOpen
  \bibfield  {author} {\bibinfo {author} {\bibfnamefont {Y.}~\bibnamefont
  {Lu}}\ and\ \bibinfo {author} {\bibfnamefont {J.}~\bibnamefont {Gao}},\
  }\href@noop {} {\bibfield  {journal} {\bibinfo  {journal} {Appl. Phys.
  Lett.}\ }\textbf {\bibinfo {volume} {101}},\ \bibinfo {pages} {043112}
  (\bibinfo {year} {2012})}\BibitemShut {NoStop}%
\bibitem [{\citenamefont {Serov}\ \emph {et~al.}(2013)\citenamefont {Serov},
  \citenamefont {Ong},\ and\ \citenamefont {Pop}}]{serov2013}%
  \BibitemOpen
  \bibfield  {author} {\bibinfo {author} {\bibfnamefont {A.~Y.}\ \bibnamefont
  {Serov}}, \bibinfo {author} {\bibfnamefont {Z.-Y.}\ \bibnamefont {Ong}}, \
  and\ \bibinfo {author} {\bibfnamefont {E.}~\bibnamefont {Pop}},\ }\href@noop
  {} {\bibfield  {journal} {\bibinfo  {journal} {Appl. Phys. Lett.}\ }\textbf
  {\bibinfo {volume} {102}},\ \bibinfo {pages} {033104} (\bibinfo {year}
  {2013})}\BibitemShut {NoStop}%
\bibitem [{\citenamefont {Aksamija}\ and\ \citenamefont
  {Knezevic}(2014)}]{aksamija2014}%
  \BibitemOpen
  \bibfield  {author} {\bibinfo {author} {\bibfnamefont {Z.}~\bibnamefont
  {Aksamija}}\ and\ \bibinfo {author} {\bibfnamefont {I.}~\bibnamefont
  {Knezevic}},\ }\href@noop {} {\bibfield  {journal} {\bibinfo  {journal}
  {Phys. Rev. B}\ }\textbf {\bibinfo {volume} {90}},\ \bibinfo {pages} {035419}
  (\bibinfo {year} {2014})}\BibitemShut {NoStop}%
\bibitem [{\citenamefont {Cai}\ \emph {et~al.}(2010)\citenamefont {Cai},
  \citenamefont {Moore}, \citenamefont {Zhu}, \citenamefont {Li}, \citenamefont
  {Chen}, \citenamefont {Shi},\ and\ \citenamefont {Ruoff}}]{cai2010}%
  \BibitemOpen
  \bibfield  {author} {\bibinfo {author} {\bibfnamefont {W.}~\bibnamefont
  {Cai}}, \bibinfo {author} {\bibfnamefont {A.~L.}\ \bibnamefont {Moore}},
  \bibinfo {author} {\bibfnamefont {Y.}~\bibnamefont {Zhu}}, \bibinfo {author}
  {\bibfnamefont {X.}~\bibnamefont {Li}}, \bibinfo {author} {\bibfnamefont
  {S.}~\bibnamefont {Chen}}, \bibinfo {author} {\bibfnamefont {L.}~\bibnamefont
  {Shi}}, \ and\ \bibinfo {author} {\bibfnamefont {R.~S.}\ \bibnamefont
  {Ruoff}},\ }\href@noop {} {\bibfield  {journal} {\bibinfo  {journal} {Nano
  Lett.}\ }\textbf {\bibinfo {volume} {10}},\ \bibinfo {pages} {1645} (\bibinfo
  {year} {2010})}\BibitemShut {NoStop}%
\bibitem [{\citenamefont {Balandin}\ \emph {et~al.}(2008)\citenamefont
  {Balandin}, \citenamefont {Ghosh}, \citenamefont {Bao}, \citenamefont
  {Calizo}, \citenamefont {Teweldebrhan}, \citenamefont {Miao},\ and\
  \citenamefont {Lau}}]{balandin2008}%
  \BibitemOpen
  \bibfield  {author} {\bibinfo {author} {\bibfnamefont {A.~A.}\ \bibnamefont
  {Balandin}}, \bibinfo {author} {\bibfnamefont {S.}~\bibnamefont {Ghosh}},
  \bibinfo {author} {\bibfnamefont {W.}~\bibnamefont {Bao}}, \bibinfo {author}
  {\bibfnamefont {I.}~\bibnamefont {Calizo}}, \bibinfo {author} {\bibfnamefont
  {D.}~\bibnamefont {Teweldebrhan}}, \bibinfo {author} {\bibfnamefont
  {F.}~\bibnamefont {Miao}}, \ and\ \bibinfo {author} {\bibfnamefont {C.~N.}\
  \bibnamefont {Lau}},\ }\href@noop {} {\bibfield  {journal} {\bibinfo
  {journal} {Nano Lett.}\ }\textbf {\bibinfo {volume} {8}},\ \bibinfo {pages}
  {902} (\bibinfo {year} {2008})}\BibitemShut {NoStop}%
\bibitem [{\citenamefont {Yasaei}\ \emph {et~al.}(2015)\citenamefont {Yasaei},
  \citenamefont {Fathizadeh}, \citenamefont {Hantehzadeh}, \citenamefont
  {Majee}, \citenamefont {El-Ghandour}, \citenamefont {Estrada}, \citenamefont
  {Foster}, \citenamefont {Aksamija}, \citenamefont {Khalili-Araghi},\ and\
  \citenamefont {Salehi-Khojin}}]{yasaei2015}%
  \BibitemOpen
  \bibfield  {author} {\bibinfo {author} {\bibfnamefont {P.}~\bibnamefont
  {Yasaei}}, \bibinfo {author} {\bibfnamefont {A.}~\bibnamefont {Fathizadeh}},
  \bibinfo {author} {\bibfnamefont {R.}~\bibnamefont {Hantehzadeh}}, \bibinfo
  {author} {\bibfnamefont {A.~K.}\ \bibnamefont {Majee}}, \bibinfo {author}
  {\bibfnamefont {A.}~\bibnamefont {El-Ghandour}}, \bibinfo {author}
  {\bibfnamefont {D.}~\bibnamefont {Estrada}}, \bibinfo {author} {\bibfnamefont
  {C.}~\bibnamefont {Foster}}, \bibinfo {author} {\bibfnamefont
  {Z.}~\bibnamefont {Aksamija}}, \bibinfo {author} {\bibfnamefont
  {F.}~\bibnamefont {Khalili-Araghi}}, \ and\ \bibinfo {author} {\bibfnamefont
  {A.}~\bibnamefont {Salehi-Khojin}},\ }\href@noop {} {\bibfield  {journal}
  {\bibinfo  {journal} {Nano Lett.}\ }\textbf {\bibinfo {volume} {15}},\
  \bibinfo {pages} {4532} (\bibinfo {year} {2015})}\BibitemShut {NoStop}%
\bibitem [{\citenamefont {Ma}\ \emph {et~al.}(2017)\citenamefont {Ma},
  \citenamefont {Liu}, \citenamefont {Wen}, \citenamefont {Gao}, \citenamefont
  {Ren}, \citenamefont {Chen}, \citenamefont {Jin}, \citenamefont {Ma},
  \citenamefont {Xu}, \citenamefont {Cheng},\ and\ \citenamefont
  {Ren}}]{ma2017}%
  \BibitemOpen
  \bibfield  {author} {\bibinfo {author} {\bibfnamefont {T.}~\bibnamefont
  {Ma}}, \bibinfo {author} {\bibfnamefont {Z.}~\bibnamefont {Liu}}, \bibinfo
  {author} {\bibfnamefont {J.}~\bibnamefont {Wen}}, \bibinfo {author}
  {\bibfnamefont {Y.}~\bibnamefont {Gao}}, \bibinfo {author} {\bibfnamefont
  {X.}~\bibnamefont {Ren}}, \bibinfo {author} {\bibfnamefont {H.}~\bibnamefont
  {Chen}}, \bibinfo {author} {\bibfnamefont {C.}~\bibnamefont {Jin}}, \bibinfo
  {author} {\bibfnamefont {X.-L.}\ \bibnamefont {Ma}}, \bibinfo {author}
  {\bibfnamefont {N.}~\bibnamefont {Xu}}, \bibinfo {author} {\bibfnamefont
  {H.-M.}\ \bibnamefont {Cheng}}, \ and\ \bibinfo {author} {\bibfnamefont
  {W.}~\bibnamefont {Ren}},\ }\href@noop {} {\bibfield  {journal} {\bibinfo
  {journal} {Nat. Commun.}\ }\textbf {\bibinfo {volume} {8}},\ \bibinfo {pages}
  {14486} (\bibinfo {year} {2017})}\BibitemShut {NoStop}%
\bibitem [{\citenamefont {Lee}\ \emph {et~al.}(2017)\citenamefont {Lee},
  \citenamefont {Kihm}, \citenamefont {Kim}, \citenamefont {Shin},
  \citenamefont {Lee}, \citenamefont {Park}, \citenamefont {Cheon},
  \citenamefont {Kwon}, \citenamefont {Lim},\ and\ \citenamefont
  {Lee}}]{lee2017}%
  \BibitemOpen
  \bibfield  {author} {\bibinfo {author} {\bibfnamefont {W.}~\bibnamefont
  {Lee}}, \bibinfo {author} {\bibfnamefont {K.~D.}\ \bibnamefont {Kihm}},
  \bibinfo {author} {\bibfnamefont {H.~G.}\ \bibnamefont {Kim}}, \bibinfo
  {author} {\bibfnamefont {S.}~\bibnamefont {Shin}}, \bibinfo {author}
  {\bibfnamefont {C.}~\bibnamefont {Lee}}, \bibinfo {author} {\bibfnamefont
  {J.~S.}\ \bibnamefont {Park}}, \bibinfo {author} {\bibfnamefont
  {S.}~\bibnamefont {Cheon}}, \bibinfo {author} {\bibfnamefont {O.~M.}\
  \bibnamefont {Kwon}}, \bibinfo {author} {\bibfnamefont {G.}~\bibnamefont
  {Lim}}, \ and\ \bibinfo {author} {\bibfnamefont {W.}~\bibnamefont {Lee}},\
  }\href@noop {} {\bibfield  {journal} {\bibinfo  {journal} {Nano Lett.}\
  }\textbf {\bibinfo {volume} {17}},\ \bibinfo {pages} {2361} (\bibinfo {year}
  {2017})}\BibitemShut {NoStop}%
\bibitem [{\citenamefont {Hirvonen}\ \emph {et~al.}(2016)\citenamefont
  {Hirvonen}, \citenamefont {Ervasti}, \citenamefont {Fan}, \citenamefont
  {Jalalvand}, \citenamefont {Seymour}, \citenamefont {Allaei}, \citenamefont
  {Provatas}, \citenamefont {Harju}, \citenamefont {Elder},\ and\ \citenamefont
  {Ala-Nissila}}]{hirvonen2016}%
  \BibitemOpen
  \bibfield  {author} {\bibinfo {author} {\bibfnamefont {P.}~\bibnamefont
  {Hirvonen}}, \bibinfo {author} {\bibfnamefont {M.~M.}\ \bibnamefont
  {Ervasti}}, \bibinfo {author} {\bibfnamefont {Z.}~\bibnamefont {Fan}},
  \bibinfo {author} {\bibfnamefont {M.}~\bibnamefont {Jalalvand}}, \bibinfo
  {author} {\bibfnamefont {M.}~\bibnamefont {Seymour}}, \bibinfo {author}
  {\bibfnamefont {S.~M.~V.}\ \bibnamefont {Allaei}}, \bibinfo {author}
  {\bibfnamefont {N.}~\bibnamefont {Provatas}}, \bibinfo {author}
  {\bibfnamefont {A.}~\bibnamefont {Harju}}, \bibinfo {author} {\bibfnamefont
  {K.~R.}\ \bibnamefont {Elder}}, \ and\ \bibinfo {author} {\bibfnamefont
  {T.}~\bibnamefont {Ala-Nissila}},\ }\href@noop {} {\bibfield  {journal}
  {\bibinfo  {journal} {Phys. Rev. B}\ }\textbf {\bibinfo {volume} {94}},\
  \bibinfo {pages} {035414} (\bibinfo {year} {2016})}\BibitemShut {NoStop}%
\bibitem [{\citenamefont {Hirvonen}\ \emph {et~al.}(2017)\citenamefont
  {Hirvonen}, \citenamefont {Fan}, \citenamefont {Ervasti}, \citenamefont
  {Harju}, \citenamefont {Elder},\ and\ \citenamefont
  {Ala-Nissila}}]{hirvonen2017}%
  \BibitemOpen
  \bibfield  {author} {\bibinfo {author} {\bibfnamefont {P.}~\bibnamefont
  {Hirvonen}}, \bibinfo {author} {\bibfnamefont {Z.}~\bibnamefont {Fan}},
  \bibinfo {author} {\bibfnamefont {M.~M.}\ \bibnamefont {Ervasti}}, \bibinfo
  {author} {\bibfnamefont {A.}~\bibnamefont {Harju}}, \bibinfo {author}
  {\bibfnamefont {K.~R.}\ \bibnamefont {Elder}}, \ and\ \bibinfo {author}
  {\bibfnamefont {T.}~\bibnamefont {Ala-Nissila}},\ }\href@noop {} {\bibfield
  {journal} {\bibinfo  {journal} {Scientific Reports}\ }\textbf {\bibinfo
  {volume} {7}},\ \bibinfo {pages} {4754} (\bibinfo {year} {2017})}\BibitemShut
  {NoStop}%
\bibitem [{\citenamefont {Elder}\ \emph {et~al.}(2002)\citenamefont {Elder},
  \citenamefont {Katakowski}, \citenamefont {Haataja},\ and\ \citenamefont
  {Grant}}]{elder2002}%
  \BibitemOpen
  \bibfield  {author} {\bibinfo {author} {\bibfnamefont {K.~R.}\ \bibnamefont
  {Elder}}, \bibinfo {author} {\bibfnamefont {M.}~\bibnamefont {Katakowski}},
  \bibinfo {author} {\bibfnamefont {M.}~\bibnamefont {Haataja}}, \ and\
  \bibinfo {author} {\bibfnamefont {M.}~\bibnamefont {Grant}},\ }\href@noop {}
  {\bibfield  {journal} {\bibinfo  {journal} {Phys. Rev. Lett.}\ }\textbf
  {\bibinfo {volume} {88}},\ \bibinfo {pages} {245701} (\bibinfo {year}
  {2002})}\BibitemShut {NoStop}%
\bibitem [{\citenamefont {Elder}\ and\ \citenamefont
  {Grant}(2004)}]{elder2004}%
  \BibitemOpen
  \bibfield  {author} {\bibinfo {author} {\bibfnamefont {K.~R.}\ \bibnamefont
  {Elder}}\ and\ \bibinfo {author} {\bibfnamefont {M.}~\bibnamefont {Grant}},\
  }\href@noop {} {\bibfield  {journal} {\bibinfo  {journal} {Phys. Rev. E}\
  }\textbf {\bibinfo {volume} {70}},\ \bibinfo {pages} {051605} (\bibinfo
  {year} {2004})}\BibitemShut {NoStop}%
\bibitem [{\citenamefont {Goldenfeld}\ \emph {et~al.}(2005)\citenamefont
  {Goldenfeld}, \citenamefont {Athreya},\ and\ \citenamefont
  {Dantzig}}]{goldenfeld2005}%
  \BibitemOpen
  \bibfield  {author} {\bibinfo {author} {\bibfnamefont {N.}~\bibnamefont
  {Goldenfeld}}, \bibinfo {author} {\bibfnamefont {B.~P.}\ \bibnamefont
  {Athreya}}, \ and\ \bibinfo {author} {\bibfnamefont {J.~A.}\ \bibnamefont
  {Dantzig}},\ }\href@noop {} {\bibfield  {journal} {\bibinfo  {journal} {Phys.
  Rev. E}\ }\textbf {\bibinfo {volume} {72}},\ \bibinfo {pages} {020601}
  (\bibinfo {year} {2005})}\BibitemShut {NoStop}%
\bibitem [{\citenamefont {Mkhonta}\ \emph {et~al.}(2013)\citenamefont
  {Mkhonta}, \citenamefont {Elder},\ and\ \citenamefont {Huang}}]{mkhonta2013}%
  \BibitemOpen
  \bibfield  {author} {\bibinfo {author} {\bibfnamefont {S.~K.}\ \bibnamefont
  {Mkhonta}}, \bibinfo {author} {\bibfnamefont {K.~R.}\ \bibnamefont {Elder}},
  \ and\ \bibinfo {author} {\bibfnamefont {Z.-F.}\ \bibnamefont {Huang}},\
  }\href@noop {} {\bibfield  {journal} {\bibinfo  {journal} {Phys. Rev. Lett.}\
  }\textbf {\bibinfo {volume} {111}},\ \bibinfo {pages} {035501} (\bibinfo
  {year} {2013})}\BibitemShut {NoStop}%
\bibitem [{\citenamefont {Zhang}\ \emph {et~al.}(2014)\citenamefont {Zhang},
  \citenamefont {Li},\ and\ \citenamefont {Gao}}]{zhang2014}%
  \BibitemOpen
  \bibfield  {author} {\bibinfo {author} {\bibfnamefont {T.}~\bibnamefont
  {Zhang}}, \bibinfo {author} {\bibfnamefont {X.}~\bibnamefont {Li}}, \ and\
  \bibinfo {author} {\bibfnamefont {H.}~\bibnamefont {Gao}},\ }\href@noop {}
  {\bibfield  {journal} {\bibinfo  {journal} {Extreme Mechanics Letters}\
  }\textbf {\bibinfo {volume} {1}},\ \bibinfo {pages} {3} (\bibinfo {year}
  {2014})}\BibitemShut {NoStop}%
\bibitem [{\citenamefont {Seymour}\ and\ \citenamefont
  {Provatas}(2016)}]{seymour2016}%
  \BibitemOpen
  \bibfield  {author} {\bibinfo {author} {\bibfnamefont {M.}~\bibnamefont
  {Seymour}}\ and\ \bibinfo {author} {\bibfnamefont {N.}~\bibnamefont
  {Provatas}},\ }\href@noop {} {\bibfield  {journal} {\bibinfo  {journal}
  {Phys. Rev. B}\ }\textbf {\bibinfo {volume} {93}},\ \bibinfo {pages} {035447}
  (\bibinfo {year} {2016})}\BibitemShut {NoStop}%
\bibitem [{\citenamefont {Fan}\ \emph {et~al.}(2013)\citenamefont {Fan},
  \citenamefont {Siro},\ and\ \citenamefont {Harju}}]{fan2013}%
  \BibitemOpen
  \bibfield  {author} {\bibinfo {author} {\bibfnamefont {Z.}~\bibnamefont
  {Fan}}, \bibinfo {author} {\bibfnamefont {T.}~\bibnamefont {Siro}}, \ and\
  \bibinfo {author} {\bibfnamefont {A.}~\bibnamefont {Harju}},\ }\href@noop {}
  {\bibfield  {journal} {\bibinfo  {journal} {Comput. Phys. Commun.}\ }\textbf
  {\bibinfo {volume} {184}},\ \bibinfo {pages} {1414} (\bibinfo {year}
  {2013})}\BibitemShut {NoStop}%
\bibitem [{\citenamefont {Fan}\ \emph {et~al.}(2015)\citenamefont {Fan},
  \citenamefont {Pereira}, \citenamefont {Wang}, \citenamefont {Zheng},
  \citenamefont {Donadio},\ and\ \citenamefont {Harju}}]{fan2015}%
  \BibitemOpen
  \bibfield  {author} {\bibinfo {author} {\bibfnamefont {Z.}~\bibnamefont
  {Fan}}, \bibinfo {author} {\bibfnamefont {L.~F.~C.}\ \bibnamefont {Pereira}},
  \bibinfo {author} {\bibfnamefont {H.-Q.}\ \bibnamefont {Wang}}, \bibinfo
  {author} {\bibfnamefont {J.-C.}\ \bibnamefont {Zheng}}, \bibinfo {author}
  {\bibfnamefont {D.}~\bibnamefont {Donadio}}, \ and\ \bibinfo {author}
  {\bibfnamefont {A.}~\bibnamefont {Harju}},\ }\href@noop {} {\bibfield
  {journal} {\bibinfo  {journal} {Phys. Rev. B}\ }\textbf {\bibinfo {volume}
  {92}},\ \bibinfo {pages} {094301} (\bibinfo {year} {2015})}\BibitemShut
  {NoStop}%
\bibitem [{\citenamefont {Fan}\ \emph {et~al.}(2017{\natexlab{a}})\citenamefont
  {Fan}, \citenamefont {Chen}, \citenamefont {Vierimaa},\ and\ \citenamefont
  {Harju}}]{fan2017cpc}%
  \BibitemOpen
  \bibfield  {author} {\bibinfo {author} {\bibfnamefont {Z.}~\bibnamefont
  {Fan}}, \bibinfo {author} {\bibfnamefont {W.}~\bibnamefont {Chen}}, \bibinfo
  {author} {\bibfnamefont {V.}~\bibnamefont {Vierimaa}}, \ and\ \bibinfo
  {author} {\bibfnamefont {A.}~\bibnamefont {Harju}},\ }\href@noop {}
  {\bibfield  {journal} {\bibinfo  {journal} {Comput. Phys. Commun.}\ }\textbf
  {\bibinfo {volume} {218}},\ \bibinfo {pages} {10} (\bibinfo {year}
  {2017}{\natexlab{a}})}\BibitemShut {NoStop}%
\bibitem [{gpu()}]{gpumd}%
  \BibitemOpen
  \href {https://github.com/brucefan1983/GPUMD} {}\bibinfo {note} {Accessed:
  2017-07-31}\BibitemShut {NoStop}%
\bibitem [{\citenamefont {Green}(1954)}]{green1954}%
  \BibitemOpen
  \bibfield  {author} {\bibinfo {author} {\bibfnamefont {M.~S.}\ \bibnamefont
  {Green}},\ }\href@noop {} {\bibfield  {journal} {\bibinfo  {journal} {J.
  Chem. Phys.}\ }\textbf {\bibinfo {volume} {22}},\ \bibinfo {pages} {398}
  (\bibinfo {year} {1954})}\BibitemShut {NoStop}%
\bibitem [{\citenamefont {Kubo}(1957)}]{kubo1957}%
  \BibitemOpen
  \bibfield  {author} {\bibinfo {author} {\bibfnamefont {R.}~\bibnamefont
  {Kubo}},\ }\href@noop {} {\bibfield  {journal} {\bibinfo  {journal} {J. Phys.
  Soc. Jpn.}\ }\textbf {\bibinfo {volume} {12}},\ \bibinfo {pages} {570}
  (\bibinfo {year} {1957})}\BibitemShut {NoStop}%
\bibitem [{\citenamefont {Fan}\ \emph {et~al.}(2017{\natexlab{b}})\citenamefont
  {Fan}, \citenamefont {Pereira}, \citenamefont {Hirvonen}, \citenamefont
  {Ervasti}, \citenamefont {Elder}, \citenamefont {Donadio}, \citenamefont
  {Ala-Nissila},\ and\ \citenamefont {Harju}}]{fan2017prb}%
  \BibitemOpen
  \bibfield  {author} {\bibinfo {author} {\bibfnamefont {Z.}~\bibnamefont
  {Fan}}, \bibinfo {author} {\bibfnamefont {L.~F.~C.}\ \bibnamefont {Pereira}},
  \bibinfo {author} {\bibfnamefont {P.}~\bibnamefont {Hirvonen}}, \bibinfo
  {author} {\bibfnamefont {M.~M.}\ \bibnamefont {Ervasti}}, \bibinfo {author}
  {\bibfnamefont {K.~R.}\ \bibnamefont {Elder}}, \bibinfo {author}
  {\bibfnamefont {D.}~\bibnamefont {Donadio}}, \bibinfo {author} {\bibfnamefont
  {T.}~\bibnamefont {Ala-Nissila}}, \ and\ \bibinfo {author} {\bibfnamefont
  {A.}~\bibnamefont {Harju}},\ }\href@noop {} {\bibfield  {journal} {\bibinfo
  {journal} {Phys. Rev. B}\ }\textbf {\bibinfo {volume} {95}},\ \bibinfo
  {pages} {144309} (\bibinfo {year} {2017}{\natexlab{b}})}\BibitemShut
  {NoStop}%
\bibitem [{\citenamefont {Nan}\ \emph {et~al.}(1997)\citenamefont {Nan},
  \citenamefont {Birringer}, \citenamefont {Clarke},\ and\ \citenamefont
  {Gleiter}}]{nan1997}%
  \BibitemOpen
  \bibfield  {author} {\bibinfo {author} {\bibfnamefont {C.-W.}\ \bibnamefont
  {Nan}}, \bibinfo {author} {\bibfnamefont {R.}~\bibnamefont {Birringer}},
  \bibinfo {author} {\bibfnamefont {D.~R.}\ \bibnamefont {Clarke}}, \ and\
  \bibinfo {author} {\bibfnamefont {H.}~\bibnamefont {Gleiter}},\ }\href@noop
  {} {\bibfield  {journal} {\bibinfo  {journal} {J. Appl. Phys.}\ }\textbf
  {\bibinfo {volume} {81}},\ \bibinfo {pages} {6692} (\bibinfo {year}
  {1997})}\BibitemShut {NoStop}%
\bibitem [{lam()}]{lammps}%
  \BibitemOpen
  \href {http://lammps.sandia.gov/doc/compute_heat_flux.html} {}\bibinfo {note}
  {Accessed: 2017-06-28}\BibitemShut {NoStop}%
\bibitem [{\citenamefont {Lampin}\ \emph {et~al.}(2013)\citenamefont {Lampin},
  \citenamefont {Palla}, \citenamefont {Francioso},\ and\ \citenamefont
  {Cleri}}]{lampin2013}%
  \BibitemOpen
  \bibfield  {author} {\bibinfo {author} {\bibfnamefont {E.}~\bibnamefont
  {Lampin}}, \bibinfo {author} {\bibfnamefont {P.~L.}\ \bibnamefont {Palla}},
  \bibinfo {author} {\bibfnamefont {P.-A.}\ \bibnamefont {Francioso}}, \ and\
  \bibinfo {author} {\bibfnamefont {F.}~\bibnamefont {Cleri}},\ }\href@noop {}
  {\bibfield  {journal} {\bibinfo  {journal} {J. Appl. Phys.}\ }\textbf
  {\bibinfo {volume} {114}},\ \bibinfo {pages} {033525} (\bibinfo {year}
  {2013})}\BibitemShut {NoStop}%
\bibitem [{\citenamefont {Melis}\ \emph {et~al.}(2014)\citenamefont {Melis},
  \citenamefont {Dettori}, \citenamefont {Vandermeulen},\ and\ \citenamefont
  {Colombo}}]{melis2014}%
  \BibitemOpen
  \bibfield  {author} {\bibinfo {author} {\bibfnamefont {C.}~\bibnamefont
  {Melis}}, \bibinfo {author} {\bibfnamefont {R.}~\bibnamefont {Dettori}},
  \bibinfo {author} {\bibfnamefont {S.}~\bibnamefont {Vandermeulen}}, \ and\
  \bibinfo {author} {\bibfnamefont {L.}~\bibnamefont {Colombo}},\ }\href@noop
  {} {\bibfield  {journal} {\bibinfo  {journal} {Eur. Phys. J. B}\ }\textbf
  {\bibinfo {volume} {87}},\ \bibinfo {pages} {96} (\bibinfo {year}
  {2014})}\BibitemShut {NoStop}%
\bibitem [{\citenamefont {Pop}\ \emph {et~al.}(2012)\citenamefont {Pop},
  \citenamefont {Varshney},\ and\ \citenamefont {Roy}}]{pop2012}%
  \BibitemOpen
  \bibfield  {author} {\bibinfo {author} {\bibfnamefont {E.}~\bibnamefont
  {Pop}}, \bibinfo {author} {\bibfnamefont {V.}~\bibnamefont {Varshney}}, \
  and\ \bibinfo {author} {\bibfnamefont {A.~K.}\ \bibnamefont {Roy}},\
  }\href@noop {} {\bibfield  {journal} {\bibinfo  {journal} {MRS Bulletin}\
  }\textbf {\bibinfo {volume} {37}},\ \bibinfo {pages} {1273} (\bibinfo {year}
  {2012})}\BibitemShut {NoStop}%
\bibitem [{\citenamefont {S\"a\"askilahti}\ \emph {et~al.}(2016)\citenamefont
  {S\"a\"askilahti}, \citenamefont {Oksanen}, \citenamefont {Tulkki},
  \citenamefont {McGaughey},\ and\ \citenamefont {Volz}}]{saaskilahti2016}%
  \BibitemOpen
  \bibfield  {author} {\bibinfo {author} {\bibfnamefont {K.}~\bibnamefont
  {S\"a\"askilahti}}, \bibinfo {author} {\bibfnamefont {J.}~\bibnamefont
  {Oksanen}}, \bibinfo {author} {\bibfnamefont {J.}~\bibnamefont {Tulkki}},
  \bibinfo {author} {\bibfnamefont {A.~J.~H.}\ \bibnamefont {McGaughey}}, \
  and\ \bibinfo {author} {\bibfnamefont {S.}~\bibnamefont {Volz}},\ }\href@noop
  {} {\bibfield  {journal} {\bibinfo  {journal} {AIP Advances}\ }\textbf
  {\bibinfo {volume} {6}},\ \bibinfo {pages} {121904} (\bibinfo {year}
  {2016})}\BibitemShut {NoStop}%
\bibitem [{\citenamefont {Turney}\ \emph {et~al.}(2009)\citenamefont {Turney},
  \citenamefont {McGaughey},\ and\ \citenamefont {Amon}}]{turney2009}%
  \BibitemOpen
  \bibfield  {author} {\bibinfo {author} {\bibfnamefont {J.~E.}\ \bibnamefont
  {Turney}}, \bibinfo {author} {\bibfnamefont {A.~J.~H.}\ \bibnamefont
  {McGaughey}}, \ and\ \bibinfo {author} {\bibfnamefont {C.~H.}\ \bibnamefont
  {Amon}},\ }\href@noop {} {\bibfield  {journal} {\bibinfo  {journal} {Phys.
  Rev. B}\ }\textbf {\bibinfo {volume} {79}},\ \bibinfo {pages} {224305}
  (\bibinfo {year} {2009})}\BibitemShut {NoStop}%
\bibitem [{\citenamefont {Singh}\ \emph {et~al.}(2011)\citenamefont {Singh},
  \citenamefont {Murthy},\ and\ \citenamefont {Fisher}}]{singh2011}%
  \BibitemOpen
  \bibfield  {author} {\bibinfo {author} {\bibfnamefont {D.}~\bibnamefont
  {Singh}}, \bibinfo {author} {\bibfnamefont {J.~Y.}\ \bibnamefont {Murthy}}, \
  and\ \bibinfo {author} {\bibfnamefont {T.~S.}\ \bibnamefont {Fisher}},\
  }\href@noop {} {\bibfield  {journal} {\bibinfo  {journal} {J. Appl. Phys.}\
  }\textbf {\bibinfo {volume} {110}},\ \bibinfo {pages} {113510} (\bibinfo
  {year} {2011})}\BibitemShut {NoStop}%
\bibitem [{\citenamefont {Lindsay}\ \emph {et~al.}(2010)\citenamefont
  {Lindsay}, \citenamefont {Broido},\ and\ \citenamefont
  {Mingo}}]{lindsay2010prb}%
  \BibitemOpen
  \bibfield  {author} {\bibinfo {author} {\bibfnamefont {L.}~\bibnamefont
  {Lindsay}}, \bibinfo {author} {\bibfnamefont {D.~A.}\ \bibnamefont {Broido}},
  \ and\ \bibinfo {author} {\bibfnamefont {N.}~\bibnamefont {Mingo}},\
  }\href@noop {} {\bibfield  {journal} {\bibinfo  {journal} {Phys. Rev. B}\
  }\textbf {\bibinfo {volume} {82}},\ \bibinfo {pages} {115427} (\bibinfo
  {year} {2010})}\BibitemShut {NoStop}%
\bibitem [{\citenamefont {Aksamija}\ and\ \citenamefont
  {Knezevic}(2011)}]{aksamija2011apl}%
  \BibitemOpen
  \bibfield  {author} {\bibinfo {author} {\bibfnamefont {Z.}~\bibnamefont
  {Aksamija}}\ and\ \bibinfo {author} {\bibfnamefont {I.}~\bibnamefont
  {Knezevic}},\ }\href@noop {} {\bibfield  {journal} {\bibinfo  {journal}
  {Applied Physics Letters}\ }\textbf {\bibinfo {volume} {98}},\ \bibinfo
  {pages} {141919} (\bibinfo {year} {2011})}\BibitemShut {NoStop}%
\bibitem [{\citenamefont {Majee}\ and\ \citenamefont
  {Aksamija}(2016)}]{majee2016prb}%
  \BibitemOpen
  \bibfield  {author} {\bibinfo {author} {\bibfnamefont {A.~K.}\ \bibnamefont
  {Majee}}\ and\ \bibinfo {author} {\bibfnamefont {Z.}~\bibnamefont
  {Aksamija}},\ }\href@noop {} {\bibfield  {journal} {\bibinfo  {journal}
  {Phys. Rev. B}\ }\textbf {\bibinfo {volume} {93}},\ \bibinfo {pages} {235423}
  (\bibinfo {year} {2016})}\BibitemShut {NoStop}%
\bibitem [{\citenamefont {Kuang}\ \emph {et~al.}(2016)\citenamefont {Kuang},
  \citenamefont {Lindsay}, \citenamefont {Shi}, \citenamefont {Wang},\ and\
  \citenamefont {Huang}}]{kuang2016}%
  \BibitemOpen
  \bibfield  {author} {\bibinfo {author} {\bibfnamefont {Y.}~\bibnamefont
  {Kuang}}, \bibinfo {author} {\bibfnamefont {L.}~\bibnamefont {Lindsay}},
  \bibinfo {author} {\bibfnamefont {S.}~\bibnamefont {Shi}}, \bibinfo {author}
  {\bibfnamefont {X.}~\bibnamefont {Wang}}, \ and\ \bibinfo {author}
  {\bibfnamefont {B.}~\bibnamefont {Huang}},\ }\href@noop {} {\bibfield
  {journal} {\bibinfo  {journal} {Int. J. Heat Mass Tran.}\ }\textbf {\bibinfo
  {volume} {101}},\ \bibinfo {pages} {772} (\bibinfo {year}
  {2016})}\BibitemShut {NoStop}%
\bibitem [{\citenamefont {Backofen}\ \emph {et~al.}(2014)\citenamefont
  {Backofen}, \citenamefont {Barmak}, \citenamefont {Elder},\ and\
  \citenamefont {Voigt}}]{backofen2014}%
  \BibitemOpen
  \bibfield  {author} {\bibinfo {author} {\bibfnamefont {R.}~\bibnamefont
  {Backofen}}, \bibinfo {author} {\bibfnamefont {K.}~\bibnamefont {Barmak}},
  \bibinfo {author} {\bibfnamefont {K.~R.}\ \bibnamefont {Elder}}, \ and\
  \bibinfo {author} {\bibfnamefont {A.}~\bibnamefont {Voigt}},\ }\href@noop {}
  {\bibfield  {journal} {\bibinfo  {journal} {Acta Materialia}\ }\textbf
  {\bibinfo {volume} {64}},\ \bibinfo {pages} {72} (\bibinfo {year}
  {2014})}\BibitemShut {NoStop}%
\bibitem [{\citenamefont {Tersoff}(1989)}]{tersoff1989}%
  \BibitemOpen
  \bibfield  {author} {\bibinfo {author} {\bibfnamefont {J.}~\bibnamefont
  {Tersoff}},\ }\href@noop {} {\bibfield  {journal} {\bibinfo  {journal} {Phys.
  Rev. B}\ }\textbf {\bibinfo {volume} {39}},\ \bibinfo {pages} {5566}
  (\bibinfo {year} {1989})}\BibitemShut {NoStop}%
\bibitem [{\citenamefont {Lindsay}\ and\ \citenamefont
  {Broido}(2010)}]{lindsay2010}%
  \BibitemOpen
  \bibfield  {author} {\bibinfo {author} {\bibfnamefont {L.}~\bibnamefont
  {Lindsay}}\ and\ \bibinfo {author} {\bibfnamefont {D.~A.}\ \bibnamefont
  {Broido}},\ }\href@noop {} {\bibfield  {journal} {\bibinfo  {journal} {Phys.
  Rev. B}\ }\textbf {\bibinfo {volume} {81}},\ \bibinfo {pages} {205441}
  (\bibinfo {year} {2010})}\BibitemShut {NoStop}%
\bibitem [{\citenamefont {Swope}\ \emph {et~al.}(1982)\citenamefont {Swope},
  \citenamefont {Andersen}, \citenamefont {Berens},\ and\ \citenamefont
  {Wilson}}]{swope1982}%
  \BibitemOpen
  \bibfield  {author} {\bibinfo {author} {\bibfnamefont {W.~C.}\ \bibnamefont
  {Swope}}, \bibinfo {author} {\bibfnamefont {H.~C.}\ \bibnamefont {Andersen}},
  \bibinfo {author} {\bibfnamefont {P.~H.}\ \bibnamefont {Berens}}, \ and\
  \bibinfo {author} {\bibfnamefont {K.~R.}\ \bibnamefont {Wilson}},\
  }\href@noop {} {\bibfield  {journal} {\bibinfo  {journal} {J. Chem. Phys.}\
  }\textbf {\bibinfo {volume} {76}},\ \bibinfo {pages} {637} (\bibinfo {year}
  {1982})}\BibitemShut {NoStop}%
\bibitem [{\citenamefont {Berendsen}\ \emph {et~al.}(1984)\citenamefont
  {Berendsen}, \citenamefont {Postma}, \citenamefont {van Gunsteren},
  \citenamefont {DiNola},\ and\ \citenamefont {Haak}}]{berendsen1984}%
  \BibitemOpen
  \bibfield  {author} {\bibinfo {author} {\bibfnamefont {H.~J.~C.}\
  \bibnamefont {Berendsen}}, \bibinfo {author} {\bibfnamefont {J.~P.~M.}\
  \bibnamefont {Postma}}, \bibinfo {author} {\bibfnamefont {W.~F.}\
  \bibnamefont {van Gunsteren}}, \bibinfo {author} {\bibfnamefont
  {A.}~\bibnamefont {DiNola}}, \ and\ \bibinfo {author} {\bibfnamefont {J.~R.}\
  \bibnamefont {Haak}},\ }\href@noop {} {\bibfield  {journal} {\bibinfo
  {journal} {J. Chem. Phys.}\ }\textbf {\bibinfo {volume} {81}},\ \bibinfo
  {pages} {3684} (\bibinfo {year} {1984})}\BibitemShut {NoStop}%
\bibitem [{\citenamefont {S\"{a}\"{a}skilahti}\ \emph
  {et~al.}(2014)\citenamefont {S\"{a}\"{a}skilahti}, \citenamefont {Oksanen},
  \citenamefont {Tulkki},\ and\ \citenamefont {Volz}}]{Saaskilahti2014}%
  \BibitemOpen
  \bibfield  {author} {\bibinfo {author} {\bibfnamefont {K.}~\bibnamefont
  {S\"{a}\"{a}skilahti}}, \bibinfo {author} {\bibfnamefont {J.}~\bibnamefont
  {Oksanen}}, \bibinfo {author} {\bibfnamefont {J.}~\bibnamefont {Tulkki}}, \
  and\ \bibinfo {author} {\bibfnamefont {S.}~\bibnamefont {Volz}},\ }\href@noop
  {} {\bibfield  {journal} {\bibinfo  {journal} {Phys. Rev. B}\ }\textbf
  {\bibinfo {volume} {90}},\ \bibinfo {pages} {134312} (\bibinfo {year}
  {2014})}\BibitemShut {NoStop}%
\bibitem [{\citenamefont {Zhou}\ and\ \citenamefont {Hu}(2015)}]{zhou2015}%
  \BibitemOpen
  \bibfield  {author} {\bibinfo {author} {\bibfnamefont {Y.}~\bibnamefont
  {Zhou}}\ and\ \bibinfo {author} {\bibfnamefont {M.}~\bibnamefont {Hu}},\
  }\href@noop {} {\bibfield  {journal} {\bibinfo  {journal} {Phys. Rev. B}\
  }\textbf {\bibinfo {volume} {92}},\ \bibinfo {pages} {195205} (\bibinfo
  {year} {2015})}\BibitemShut {NoStop}%
\bibitem [{\citenamefont {Nos\'{e}}(1984)}]{nose1984}%
  \BibitemOpen
  \bibfield  {author} {\bibinfo {author} {\bibfnamefont {S.}~\bibnamefont
  {Nos\'{e}}},\ }\href@noop {} {\bibfield  {journal} {\bibinfo  {journal} {J.
  Chem. Phys.}\ }\textbf {\bibinfo {volume} {81}},\ \bibinfo {pages} {511}
  (\bibinfo {year} {1984})}\BibitemShut {NoStop}%
\bibitem [{\citenamefont {Hoover}(1985)}]{hoover1985}%
  \BibitemOpen
  \bibfield  {author} {\bibinfo {author} {\bibfnamefont {W.~G.}\ \bibnamefont
  {Hoover}},\ }\href@noop {} {\bibfield  {journal} {\bibinfo  {journal} {Phys.
  Rev. A}\ }\textbf {\bibinfo {volume} {31}},\ \bibinfo {pages} {1695}
  (\bibinfo {year} {1985})}\BibitemShut {NoStop}%
\bibitem [{\citenamefont {Martyna}\ \emph {et~al.}(1992)\citenamefont
  {Martyna}, \citenamefont {Tuckerman},\ and\ \citenamefont
  {Klein}}]{martyna1992}%
  \BibitemOpen
  \bibfield  {author} {\bibinfo {author} {\bibfnamefont {G.~J.}\ \bibnamefont
  {Martyna}}, \bibinfo {author} {\bibfnamefont {M.~E.}\ \bibnamefont
  {Tuckerman}}, \ and\ \bibinfo {author} {\bibfnamefont {M.~L.}\ \bibnamefont
  {Klein}},\ }\href@noop {} {\bibfield  {journal} {\bibinfo  {journal} {J.
  Chem. Phys.}\ }\textbf {\bibinfo {volume} {97}},\ \bibinfo {pages} {2635}
  (\bibinfo {year} {1992})}\BibitemShut {NoStop}%
\end{thebibliography}%

\end{document}